\documentclass[aps,prb,twocolumn,superscriptaddress,groupedaddress,a4paper]{revtex4} 
\usepackage{graphicx}
\usepackage{dcolumn}
\usepackage{bm}
\usepackage{amssymb}
\usepackage{amsmath}
\usepackage{subfigure}
\usepackage{color}
\newcommand{\Eg}{$E_{g}$}
\newcommand{\DeltaSO}{$\Delta_{\scalebox{0.6}{\textrm{SO}}}$}

\begin{document}

\widetext
\title{Investigation of the anisotropic electron g factor as a probe of the electronic structure of GaBi$_{x}$As$_{1-x}$/GaAs epilayers}


\author{Christopher A. Broderick}
\email{c.broderick@umail.ucc.ie} 
\affiliation{Tyndall National Institute, Lee Maltings, Dyke Parade, Cork, Ireland}
\affiliation{Department of Physics, University College Cork, Cork, Ireland} 

\author{Simone Mazzucato}
\email{mazzucat@insa-toulouse.fr} 
\affiliation{LPCNO, INSA-UPS-CNRS, 135 Avenue de Rangueil, F-31400 Toulouse, France}

\author{H\'{e}l\`{e}ne Carr\`{e}re}
\affiliation{LPCNO, INSA-UPS-CNRS, 135 Avenue de Rangueil, F-31400 Toulouse, France}

\author{Thierry Amand}
\affiliation{LPCNO, INSA-UPS-CNRS, 135 Avenue de Rangueil, F-31400 Toulouse, France}

\author{Hejer Makhloufi}
\affiliation{LAAS-CNRS, 7 Avenue du Colonel Roche, F-31400 Toulouse, France}

\author{Alexandre Arnoult}
\affiliation{LAAS-CNRS, 7 Avenue du Colonel Roche, F-31400 Toulouse, France}

\author{Chantal Fontaine}
\affiliation{LAAS-CNRS, 7 Avenue du Colonel Roche, F-31400 Toulouse, France}

\author{Omer Donmez}
\affiliation{Department of Physics, Faculty of Science, Istanbul University, Vezneciler, Istanbul 34134, Turkey}

\author{Ay\c{s}e Erol}
\affiliation{Department of Physics, Faculty of Science, Istanbul University, Vezneciler, Istanbul 34134, Turkey}

\author{Muhammad Usman}
\affiliation{Centre for Quantum Computation and Communication Technology, School of Physics, University of Melbourne, Parkville, Victoria 3010, Australia}

\author{Eoin P. O'Reilly}
\affiliation{Tyndall National Institute, Lee Maltings, Dyke Parade, Cork, Ireland}
\affiliation{Department of Physics, University College Cork, Cork, Ireland}

\author{Xavier Marie}
\affiliation{LPCNO, INSA-UPS-CNRS, 135 Avenue de Rangueil, F-31400 Toulouse, France}

\date{\today}


\begin{abstract}

The electron Land\'{e} g factor ($g^{*}$) is investigated both experimentally and theoretically in a series of GaBi$_{x}$As$_{1-x}$/GaAs strained epitaxial layers, for bismuth compositions up to $x = 3.8$\%. We measure $g^{*}$ via time-resolved photoluminescence spectroscopy, which we use to analyze the spin quantum beats in the polarization of the photoluminescence in the presence of an externally applied magnetic field. The experimental measurements are compared directly to atomistic tight-binding calculations on large supercells, which allows us to explicitly account for alloy disorder effects. We demonstrate that the magnitude of $g^{*}$ increases strongly with increasing Bi composition $x$ and, based on the agreement between the theoretical calculations and experimental measurements, elucidate the underlying causes of the observed variation of $g^{*}$. By performing measurements in which the orientation of the applied magnetic field is changed, we further demonstrate that $g^{*}$ is strongly anisotropic. We quantify the observed variation of $g^{*}$ with $x$, and its anisotropy, in terms of a combination of epitaxial strain and Bi-induced hybridization of valence states due to alloy disorder, which strongly perturbs the electronic structure.

\end{abstract}


\maketitle


\section{Introduction}
\label{sec:introduction}

The incorporation of small concentrations of bismuth (Bi) into GaAs to form the highly-mismatched dilute bismide alloy GaBi$_{x}$As$_{1-x}$ results in a rapid reduction of the band gap (\Eg) with increasing Bi composition $x$, of 70 -- 80 meV per \% Bi replacing As in the alloy at low $x$. \cite{Tixier_APL_2005,Alberi_PRB_2007} This decrease in \Eg\, is accompanied by a strong increase of the spin-orbit-splitting energy (\DeltaSO), \cite{Fluegel_PRL_2006} which leads to a band structure in which $\Delta_{\scalebox{0.6}{\textrm{SO}}} > E_{g}$ for $x \gtrsim 10$\%. \cite{Usman_PRB_2011,Batool_JAP_2012} This band structure condition makes dilute bismide alloys an attractive candidate material system for applications in GaAs-based semiconductor lasers operating at mid-infrared wavelengths due to the possibility to suppress the dominant non-radiative recombination pathway. \cite{Sweeney_JAP_2013} Moreover, the increased spin-orbit coupling in dilute bismide alloys could lead to the development of spintronic devices based on the Rashba spin-orbit interaction requiring shorter channel lengths or reduced modulation electric field strengths compared to existing GaAs-based devices. \cite{Datta_APL_1990} As a result of this and other potential applications, research interest in dilute bismide alloys has been steadily increasing in recent years.

As a result of the large differences in size and chemical properties between Bi atoms and the As atoms they replace, Bi acts as an isovalent impurity in GaAs which strongly perturbs the valence band (VB) structure. An explanation of the observed strong, composition-dependent bowing of \Eg\, and \DeltaSO\, in GaBi$_{x}$As$_{1-x}$ has been proposed \cite{Alberi_PRB_2007,Usman_PRB_2011,Broderick_SST_2013} in the form of a band anti-crossing interaction between the extended VB edge states of the GaAs host matrix, and highly localized Bi-related impurity states which are resonant with the GaAs VB. The situation in disordered GaBi$_{x}$As$_{1-x}$ alloys is more complicated, where the reduction in symmetry brought about through alloy disorder has significant effects on the electronic properties. \cite{Usman_PRB_2011,Usman_PRB_2013}

The Land\'{e} g factor of conduction electrons ($g^{*}$) is a key parameter in semiconductor materials, describing the response of the electron spins to externally applied magnetic fields, as well as providing useful information regarding the symmetry of the electron and hole states at the $\Gamma$ point in the Brillouin zone. Therefore, in addition to providing information pertinent to evaluating the potential of specific materials for spintronic applications, analysis of the electron spin properties and g factor gives insight into the material band structure, providing valuable data for material characterization and modeling. \cite{Alberi_PRB_2007,Broderick_SST_2013,Usman_PRB_2013} In particular, $g^{*}$ is extremely sensitive to the separation in energy between the conduction band (CB) minimum and the light-hole (LH), heavy-hole (HH) and spin-split-off (SO) bands, as well as hybridization between these bands caused by, e.g., a reduction of crystal symmetry due to pseudomorphic strain. \cite{Hendorfer_SST_1991} As a result, calculation of $g^{*}$ provides a stringent test of theoretical models of the material band structure. \cite{Oestreich_IEEEJSTQE_1996}

The Larmor precession of electron spins in the presence of a magnetic field leads to quantum beating of the circular polarization of the photoluminescence (PL). Using time-resolved PL spectroscopy the frequency of these quantum beats can be determined, allowing the relevant component of $g^{*}$ to be determined directly for a given orientation of the externally applied field. \cite{LeJeune_SST_1997} As a result of the long coherence times associated with the spin quantum beats observed in the PL polarization, this technique enables highly accurate determination of $g^{*}$. \cite{Heberle_PRL_1994,Oestreich_IEEEJSTQE_1996}

Recently, in Ref.~\onlinecite{Mazzucato_APL_2013}, we experimentally studied the temperature-dependent spin dynamics of a single GaBi$_{x}$As$_{1-x}$/GaAs epilayer with $x = 2.2$\% and presented the first measurement of the transverse component of the electron g factor ($g_{\perp}^{*}$) in this emerging material system. Our measurements showed that the magnitude of $g_{\perp}^{*}$ increases strongly with increasing Bi composition $x$ in GaBi$_{x}$As$_{1-x}$, primarily due to the increase in the strength of the spin-orbit coupling with increasing $x$. In this article we significantly extend upon the analysis of Ref.~\onlinecite{Mazzucato_APL_2013} by studying a number of samples with Bi compositions ranging up to $x = 3.8$\%. This allows us to examine the evolution of $g^{*}$ with $x$, which enables us to identify general trends in the spin properties of GaBi$_{x}$As$_{1-x}$ alloys. By performing time-resolved PL measurements with different orientations of the applied magnetic field, we further demonstrate that $g^{*}$ is strongly anisotropic in GaBi$_{x}$As$_{1-x}$ epilayers.

This experimental analysis is supported by the development of a detailed theoretical model for the calculation of $g^{*}$ in GaBi$_{x}$As$_{1-x}$ and related alloys. The theoretical model is based upon a well-established, atomistic $sp^{3}s^{*}$ tight-binding Hamiltonian which we have previously applied to the study of the electronic structure of ordered \cite{Broderick_SST_2013} and disordered \cite{Usman_PRB_2011,Usman_PRB_2013} GaBi$_{x}$As$_{1-x}$ alloys, as well as GaBi$_{x}$As$_{1-x}$-based quantum well heterostructures. \cite{Usman_APL_2014} Using this model we identify and quantify the causes of the observed anisotropy and strong variation of $g^{*}$ with $x$ in GaBi$_{x}$As$_{1-x}$. Our theoretical calculations explain the measured variation of $g^{*}$ with Bi composition $x$ in terms of the effects of epitaxial strain and Bi-induced mixing of VB states due to alloy disorder. By employing large supercell calculations containing 4096 atoms we account for the effects of alloy disorder in a realistic manner and hence obtain a theoretical description of $g^{*}$ which is in very good agreement with experiment.

The remainder of this paper is organized as follows. In Section~\ref{sec:experimental_details} we describe the sample growth and characterization, as well as the details of the experimental measurements. This is followed in Section~\ref{sec:theoretical_model} by an outline of our theoretical model. We then present the results of our combined experimental and theoretical study in Section~\ref{sec:results}, beginning in Sections~\ref{sec:spectroscopy_results} and~\ref{sec:measurement_of_g} with time-integrated and time-resolved experimental results, before presenting our theoretical results in Section~\ref{sec:theoretical_results}. Finally, in Section~\ref{sec:conclusions} we conclude.


\section{Experimental details}
\label{sec:experimental_details}

We have investigated five undoped GaBi$_{x}$As$_{1-x}$ epilayers with Bi compositions up to $x = 3.8$\%, which were grown on (001)-oriented GaAs substrates by solid-source molecular beam epitaxy. Full details of the growth parameters can be found in Refs.~\onlinecite{Sarcan_NRL_2014} and~\onlinecite{Mazzucato_NRL_2014}. The Bi composition in each epilayer was determined using x-ray diffraction (XRD). Additionally, the XRD measurements indicated that the GaBi$_{x}$As$_{1-x}$ epilayers were elastically strained. The structural details of each of the epilayer samples are listed in Table~\ref{tab:sample_details}.


\begin{table}[tb!]
		\caption{\label{tab:sample_details} Structural parameters for the GaBi$_{x}$As$_{1-x}$/GaAs epilayers investigated, including layer thicknesses and compositions, as well as the expected growth uncertainties in these quantities provided by the Bruker analysis software LEPTOS.}
	\begin{ruledtabular}
		\begin{center}
			\begin{tabular}{ccc}
				\, Sample \, & Thickness (nm) \, & \, Bi composition, $x$ (\%) \, \\
				\hline
				1 & $233 \pm 1$  & $1.16 \pm 0.01$ \\
				2 & $230 \pm 10$ & $1.80 \pm 0.05$ \\
				3 & $258 \pm 1$  & $2.34 \pm 0.01$ \\
				4 & $147 \pm 1$  & $3.04 \pm 0.02$ \\
				5 & $265 \pm 1$  & $3.83 \pm 0.02$ \\
			\end{tabular}
		\end{center}
	\end{ruledtabular}
\end{table}

A combination of time-integrated and time-resolved PL spectroscopy was used to investigate the optical response and spin properties of each of the samples. Optical excitation was provided via 1.5 ps long pulses generated by a mode-locked Ti:Al$_{2}$O$_{3}$ laser with 80 MHz repetition frequency. The excitation wavelength was fixed at 795 nm and the beam was focused on a 50 $\mu$m diameter spot on the sample surface. The polarization of the exciting beam (linear or circular) was set by a combination of a linear polarizer and a quarter-wave plate, while the degree of circular polarization was obtained by passing the PL signal through another quarter-wave plate followed by a linear polarizer. The PL signals were recorded using a S1 photocathode streak camera with an overall temporal resolution of 8 ps.

For the measurement of the anisotropic electron g factor continuous external magnetic fields up to 0.73 T were applied parallel to (Voigt configuration), and at a 45$^{\circ}$ angle from (tilted configuration), the plane of the GaBi$_{x}$As$_{1-x}$ epilayer. \cite{LeJeune_SST_1997} We note that, in this range of magnetic field intensities, the binding energy of excitons is not strengthened by Landau quantization effects. \cite{Pettinari_PRB_2011} The incident laser beam was circularly polarized. In the Voigt configuration, the light propagation was along the sample growth ($z$-) direction while in the tilted configuration, due to the  high refractive index of GaBi$_{x}$As$_{1-x}$, the light propagated at an angle of approximately $\sim 12^{\circ}$ from the growth direction. In this case, the circular polarization of the exciting light was still close to 100\%, and the mean electron spin orientation was almost entirely along the $z$-axis.

The samples were mounted in a closed cycle He-cooled cryostat, and the temperature varied from 100 to 300 K. Based on an analysis of the low temperature PL \cite{Mazzucato_NRL_2014} the temperatures and excitation intensities for the experiments were chosen in order to minimize both excitonic effects and the effects of radiative recombination via localized states. \cite{Mazzucato_SST_2013,Mohmad_PSSB_2014} Therefore, all measurements of $g^{*}$ were carried out at temperatures of 100 K and above with excitation powers exceeding 10 mW, corresponding to incident photon fluxes at the sample surface of at least $2.5 \times 10^{12}$ cm$^{-2}$. 

Since the value of $g^{*}$ is strongly dependent upon the material band gap and spin-orbit-splitting energies, \cite{Hermann_PRB_1977,Hendorfer_SST_1991} photo-modulated reflectance (PR) measurements were carried out on each of the samples, which enabled both \Eg\, and \DeltaSO\, to be measured accurately. \cite{Glembocki_SS_1992,Misiewicz_OR_2000} The PR measurements were carried out in the bright configuration, in which the samples were illuminated by quasi-white light from a 50 W halogen lamp. A 10 mW beam from a 514 nm Ar-ion laser chopped at a frequency of 170 Hz was used as the modulation source. The reflected light was dispersed through a 0.5 m monochromator and detected by a thermoelectrically cooled InGaAs photodiode. The AC and DC components of the reflectance ($R$) and the relative change in the reflectance ($\frac{\Delta R}{R}$) were simultaneously acquired using a lock-in amplifier.  \cite{Donmez_NRL_2011} The PR spectra measured in this manner were then analyzed using low-field PR line-shape functions based on the Aspnes third-derivative functional form. \cite{Aspnes_SS_1973} Based on this analysis, the dependence of \Eg\, and \DeltaSO\, on the Bi composition $x$ was determined directly from the measured PR spectra.


\section{Theoretical model}
\label{sec:theoretical_model}

In this section we outline the theory of the effects of Bi incorporation on the electron g factor in GaBi$_{x}$As$_{1-x}$. The calculation of $g^{*}$ is based on an extension of the method outlined in Ref.~\onlinecite{Hendorfer_SST_1991}, which considered the effects of pseudomorphic strain on the transverse and longitudinal components of $g^{*}$ via the strain-induced hybridization of the LH and SO band edge states at $\Gamma$.

We extend this approach here to consider the hybridization of the extended states of the GaAs VB edge with Bi-related localized states, which we describe using large supercell calculations. A disordered GaBi$_{x}$As$_{1-x}$ alloy contains a large range of these localized states which are associated not only with isolated Bi atoms, but also with Bi-Bi pairs and larger clusters of Bi atoms. We have previously shown \cite{Usman_PRB_2011,Usman_PRB_2013} that these alloy disorder and hybridization effects play a crucial role in determining the details of the VB structure in GaBi$_{x}$As$_{1-x}$.

Our electronic structure calculations are based upon an atomistic $sp^{3}s^{*}$ tight-binding Hamiltonian for GaBi$_{x}$As$_{1-x}$, which includes the effects of spin-orbit coupling. In this model the on-site energies are taken to depend on the overall neighbor environment and the inter-atomic interaction matrix elements are taken to vary with the relaxed nearest-neighbor bond length $d$ as $\left( \frac{ d_{0} }{ d } \right)^{\eta}$, where $d_{0}$ is the equilibrium bond length in the equivalent binary compound and $\eta$ is a dimensionless scaling parameter (the value of which depends on the type and symmetry of the interaction). The bond angle dependence of the interaction matrix elements is represented using the two-center integrals of Slater and Koster. \cite{Slater_PR_1954} The supercell calculations were implemented using the NanoElectronic MOdeling package NEMO-3D. \cite{Klimeck_IEEETED_2007_1,Klimeck_IEEETED_2007_2} Full details of the tight-binding model can be found in Ref.~\onlinecite{Usman_PRB_2011}.

The Bi-induced mixing of VB states can be quantified by the fractional GaAs $\Gamma$ character of the zone center alloy states. For a given GaBi$_{x}$As$_{1-x}$ zone center state $\vert \psi_{j,1} \rangle$ having energy $E_{j,1}$, the fractional GaAs $\Gamma$ character is given by \cite{Usman_PRB_2011,Broderick_SST_2013}

\begin{equation}
	G_{\Gamma} ( E_{j,1} ) = \sum_{ i = 1 }^{ g ( E_{i,0} ) } \vert \langle \psi_{j,1} \vert \psi_{i,0} \rangle \vert^{2} \, ,
	\label{eq:gamma_character}
\end{equation}

\noindent
where $\vert \psi_{i,0} \rangle$ is the unperturbed GaAs zone center host matrix state having energy $E_{i,0}$ and $g ( E_{i,0} )$ is the degeneracy of $E_{i,0}$ in the unstrained bulk material. $G_{\Gamma} ( E_{j,1} )$ is therefore a direct measure of the fraction of the host matrix GaAs states at energy $E_{i,0}$ which have mixed into the alloy state $\vert \psi_{j,1} \rangle$.

The GaBi$_{x}$As$_{1-x}$ epilayers under investigation were grown in a state of biaxial compressive stress, so that an appropriate macroscopic strain must be included in the supercell calculations in order to correctly describe the epilayer electronic properties. For a pseudomorphic epilayer under biaxial stress, the longitudinal and transverse components of the strain tensor are related through Poisson's ratio $\sigma$ by \cite{Landau_elasticity}

\begin{equation}
	\epsilon_{\perp} = \frac{ - 2 \sigma }{ 1 - \sigma } \, \epsilon_{\parallel} \, ,
\end{equation}

\noindent
with the longitudinal strain in the plane of the epilayer given by

\begin{equation}
	\epsilon_{\parallel} = \frac{ a_{0} - a(x) }{ a(x) } \, ,
\end{equation}

\noindent
where $a_{0}$ and $a(x)$ are, respectively, the lattice constants of the GaAs substrate and the GaBi$_{x}$As$_{1-x}$ epilayer. The GaBi$_{x}$As$_{1-x}$ lattice constants along the $x$, $y$ and $z$ directions in the strained epilayer are then given by $a_{x} = a_{y} = a_{0}$ and $a_{z} = ( 1 + \epsilon_{\perp} ) \, a(x)$.

We interpolate the GaBi$_{x}$As$_{1-x}$ lattice and elastic constants linearly between those of the end point binary compounds \cite{Vurgaftman_JAP_2001,Ferhat_PRB_2006} in order to calculate $\epsilon_{\parallel}$ and $\epsilon_{\perp}$ for a given Bi composition $x$, and use the lattice constants $a_{x}$, $a_{y}$ and $a_{z}$ to set the initial atomic positions in the supercell. Beginning with a GaAs supercell in which the atomic positions are defined in this manner, a disordered GaBi$_{x}$As$_{1-x}$ supercell is generated by replacing As atoms by Bi atoms at randomly chosen sites on the anion sublattice. Then, in order to impose a macroscopic pseudomorphic strain on the supercell, we fix the positions of the atoms on the exterior boundaries and perform a relaxation of the positions of the atoms in the interior of the supercell using a valence force field model based on the Keating potential. \cite{Keating_PR_1966,Lazarenkova_APL_2004}

In a pseudomorphically strained epilayer of a conventional III-V alloy, the transverse ($g_{\perp}^{*}$) and longitudinal ($g_{\parallel}^{*}$) components of $g^{*}$ are given in the limit of small strain by \cite{Hendorfer_SST_1991}

\begin{eqnarray}
	\frac{ g_{\perp}^{*} }{ g_{0} } &=& 1 - \frac{ 2 m_{0} \vert P \vert^{2} }{ 3 \hbar^{2} } \left( \frac{1}{E_{B}} - \frac{1}{E_{C}} \right) \, ,	\label{eq:g_factor_hendorfer_per} \\
	\frac{ g_{\parallel}^{*} }{ g_{0} } &=& 1 - \frac{ m_{0} \vert P \vert^{2} }{ \hbar^{2} } \left( \frac{1}{E_{A}} - \frac{1}{3E_{B}} - \frac{2}{3E_{C}} \right) \, ,	\label{eq:g_factor_hendorfer_par}
\end{eqnarray}

\noindent
where $m_{0}$ and $g_{0}$ are, respectively, the free electron mass and g factor. In Eqs.~\eqref{eq:g_factor_hendorfer_per} and~\eqref{eq:g_factor_hendorfer_par} the indices $A$, $B$ and $C$ denote, respectively, the HH, LH and SO band edge states at $\Gamma$, while $E_{A}$, $E_{B}$ and $E_{C}$ denote the differences in energy between each of these states and the $\Gamma_{6c}$ states at the CB minimum. The inter-band momentum matrix element $P$ between the $s$-like $\Gamma_{6c}$ states at the CB minimum and the $p$-like $\Gamma_{8v}$ states at the VB maximum is defined by

\begin{equation}
	P = - \frac{ i \hbar }{ m_{0} } \langle s \vert \widehat{p}_{m} \vert m \rangle
\end{equation}

\noindent
where $m = x, y, z$ and $\widehat{p}_{m}$ denotes the components of the momentum operator.

In Ref.~\onlinecite{Hendorfer_SST_1991} the strain-induced hybridization between the zone center LH and SO states was calculated explicitly, leading to additional terms in Eqs.~\eqref{eq:g_factor_hendorfer_per} and~\eqref{eq:g_factor_hendorfer_par} which are proportional to the degree of mixing between them. In our analysis of pseudomorphically strained GaBi$_{x}$As$_{1-x}$ alloys here, we include such state mixing in our calculations by projecting each alloy state onto the unstrained GaAs HH, LH and SO states, as described by Eq.~\eqref{eq:gamma_character}. Using unstrained GaAs states in this way allows us to explicitly include the alloy- and strain-induced modifications to $g^{*}$ in our theoretical calculations.

In a GaBi$_{x}$As$_{1-x}$ alloy the extended VB edge states of the GaAs host matrix hybridize with a large number of alloy VB states, \cite{Broderick_SST_2013} with the details of the VB mixing depending strongly on alloy disorder. \cite{Usman_PRB_2011,Usman_PRB_2013} As discussed above, this hybridization can be quantified by the fractional GaAs $\Gamma$ character. Furthermore, the zone center alloy states $\vert \psi_{j,1} \rangle$ form a complete basis set in which the extended host matrix VB states $\vert \psi_{i,0} \rangle$ can be expanded. Therefore, in order to explicitly account for the effects of Bi- and strain-induced mixing of the HH-like VB states in a disordered GaBi$_{x}$As$_{1-x}$ supercell, we make the replacement

\begin{equation}
	\frac{ 1 }{ E_{A} } \to \sum_{j} \frac{ G_{\Gamma}^{\scalebox{0.6}{\textrm{HH}}} \left( E_{j,1} \right) }{ E_{\scalebox{0.6}{\textrm{CB}},1} - E_{j,1} } \, ,
	\label{eq:replacement}
\end{equation}

\noindent
where $E_{\scalebox{0.6}{\textrm{CB}},1}$ is the GaBi$_{x}$As$_{1-x}$ CB edge energy, the sum runs over the zone center supercell states and $G_{\Gamma}^{\scalebox{0.6}{\textrm{HH}}} \left( E_{j,1} \right)$ is the fractional GaAs HH character of the supercell VB state $\vert \psi_{j,1} \rangle$, calculated by taking $\vert \psi_{i,0} \rangle = \vert \psi_{\scalebox{0.6}{\textrm{HH}},0} \rangle$ in Eq.~\eqref{eq:gamma_character}.

Making the equivalent replacements for $E_{B}^{-1}$ and  $E_{C}^{-1}$ in Eqs.~\eqref{eq:g_factor_hendorfer_per} and~\eqref{eq:g_factor_hendorfer_par} in terms of $G_{\Gamma}^{\scalebox{0.6}{\textrm{LH}}} \left( E_{j,1} \right)$ and $G_{\Gamma}^{\scalebox{0.6}{\textrm{SO}}} \left( E_{j,1} \right)$, calculated using the unstrained GaAs LH and SO states $\vert \psi_{\scalebox{0.6}{\textrm{LH}},0} \rangle$ and $\vert \psi_{\scalebox{0.6}{\textrm{SO}},0} \rangle$ in Eq.~\eqref{eq:gamma_character}, we arrive at the following expressions for $g_{\perp}^{*}$ and $g_{\parallel}^{*}$ in a disordered, pseudomorphically strained GaBi$_{x}$As$_{1-x}$ alloy

\begin{widetext}
\begin{eqnarray}
	\frac{ g_{\perp}^{*} }{ g_{0} } &=& 1 - \frac{ 2 m_{0} \vert P \vert^{2} }{ 3 \hbar^{2} } \sum_{j} \frac{ G_{\Gamma}^{\scalebox{0.6}{\textrm{LH}}} \left( E_{j,1} \right) - G_{\Gamma}^{\scalebox{0.6}{\textrm{SO}}} \left( E_{j,1} \right) }{ E_{\scalebox{0.6}{\textrm{CB}},1} - E_{j,1} } \, , \label{eq:g_factor_per} \\
	\frac{ g_{\parallel}^{*} }{ g_{0} } &=& 1 - \frac{ 2 m_{0} \vert P \vert^{2} }{ 3 \hbar^{2} } \sum_{j} \frac{ \frac{3}{2} G_{\Gamma}^{\scalebox{0.6}{\textrm{HH}}} \left( E_{j,1} \right) - \frac{1}{2} G_{\Gamma}^{\scalebox{0.6}{\textrm{LH}}} \left( E_{j,1} \right) - G_{\Gamma}^{\scalebox{0.6}{\textrm{SO}}} \left( E_{j,1} \right) }{ E_{\scalebox{0.6}{\textrm{CB}},1} - E_{j,1} } \, , \label{eq:g_factor_par}
\end{eqnarray}
\end{widetext}

\noindent
which reduce respectively to Eqs.~\eqref{eq:g_factor_hendorfer_per} and~\eqref{eq:g_factor_hendorfer_par} for $x = 0$.

Since we are concerned with describing the variation of $g^{*}$ with Bi composition $x$, and because Bi incorporation primarily affects the VB, we have neglected the effects of the $\Gamma_{7c}$ and $\Gamma_{8c}$ CB states in our analysis of $g^{*}$. The effects of these bands on $g^{*}$ are significantly less important than the effects of the $\Gamma_{7v}$ and $\Gamma_{8v}$ VB states that we have included explicitly in our model, due to their weaker relative coupling to the $\Gamma_{6c}$ states at the CB minimum. \cite{Hermann_PRB_1977} In addition, the analysis presented here can be straightforwardly modified to include contributions from these and other remote bands.

We have previously shown \cite{Usman_PRB_2011,Broderick_PSSB_2013} that in order to accurately describe the effects of Bi-induced mixing of VB states in ordered (In)GaBi$_{x}$As$_{1-x}$ alloys, large supercells containing $\gtrsim 2000$ atoms must be considered. The use of supercells containing several thousand atoms further enables an accurate description of the effects of alloy disorder to be obtained, due to the large scope for the formation of different local configurations of group V atoms at fixed alloy composition. We therefore perform our calculations on supercells containing 4096 atoms. For a given Bi composition $x$ we consider four separate supercells, each of which contains a different statistically random spatial distribution of Bi atoms. We then obtain our theoretical values for the longitudinal and transverse components of $g^{*}$ at each Bi composition $x$ by averaging over the results obtained for each distinct supercell at that fixed composition.


\section{Results}
\label{sec:results}


\subsection{Time-integrated photoluminescence and photo-modulated reflectance}
\label{sec:spectroscopy_results}


\begin{figure}[t!]
	\includegraphics[width=1.00\columnwidth]{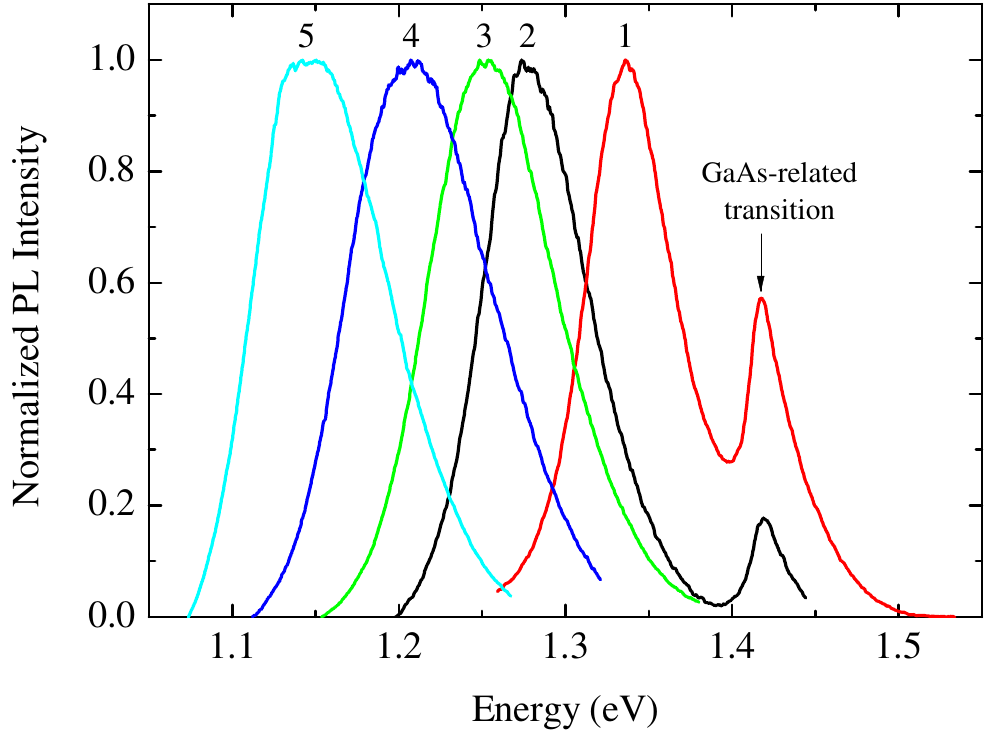}
	\caption{Time-integrated photoluminescence spectra measured at $T = 300$ K, for the five GaBi$_{x}$As$_{1-x}$ samples listed in Table~\ref{tab:sample_details}. The measured spectra for samples 1 -- 5, which have all been normalized, are shown using red, black, green, blue and light blue lines, respectively. The sharp features visible at 1.42 eV correspond to emission from the GaAs substrate.}
	\label{fig:time_integrated_pl}
\end{figure}

Steady-state PR and PL measurements have been used to investigate the optical properties of each of the GaBi$_{x}$As$_{1-x}$ epilayers. Measurements of the low-temperature PL at low excitation intensities displayed strong localization effects, characterized by an ``s-shape" variation of the PL peak energy with temperature up to approximately 100 K. \cite{Imhof_APL_2011,Imhof_PSSB_2011,Mazzucato_SST_2013,Mazzucato_NRL_2014} As stated in Section~\ref{sec:experimental_details}, in order to investigate the g factor associated with conduction electrons (as opposed to that associated with localized excitons) the PL measurements were performed at temperatures $\geq 100$ K and with sufficiently high excitation intensities to ensure at most a marginal contribution to the PL signal from localized states. \cite{Mohmad_PSSB_2014,Mazzucato_NRL_2014}


\begin{figure}[t!]
	\includegraphics[width=1.00\columnwidth]{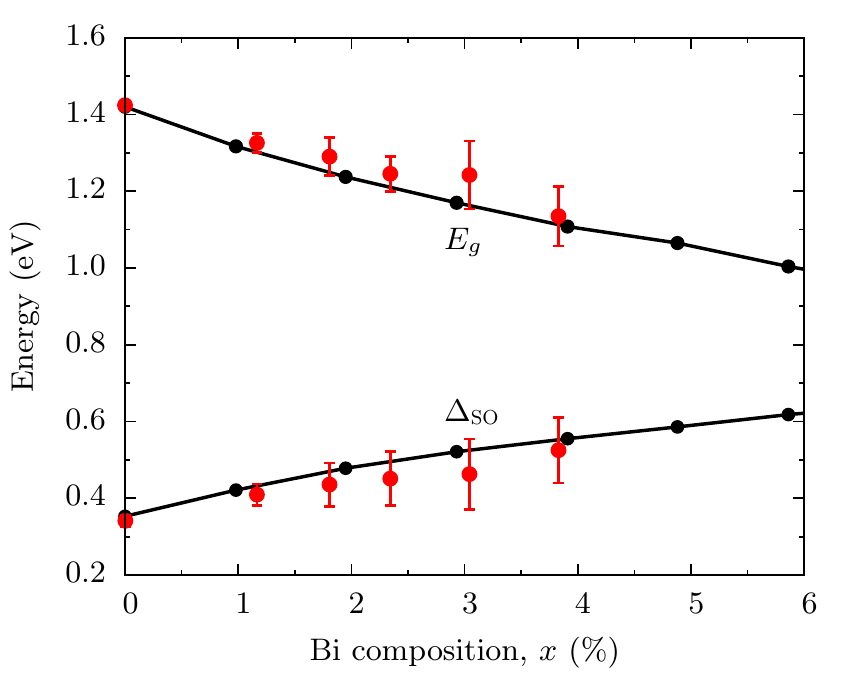}
	\caption{Measured (red closed circles) and calculated (black closed circles) variation of the band gap (\Eg) and spin-orbit-splitting energy (\protect\DeltaSO) at $T = 300$ K as a function of Bi composition $x$ for the GaBi$_{x}$As$_{1-x}$/GaAs epilayers listed in Table~\ref{tab:sample_details}. The measured values were obtained using photo-modulated reflectance spectroscopy, while the calculated values were obtained using the theoretical model outlined in Section~\ref{sec:theoretical_model}.}
	\label{fig:energy_gaps}
\end{figure}

Fig.~\ref{fig:time_integrated_pl} shows the measured PL spectra at $T = 300$ K for all five of the epilayers listed in Table~\ref{tab:sample_details}. The measured red-shift of the PL peak over the investigated composition range is approximately 75 meV per \% Bi, in agreement with previous experimental \cite{Fluegel_PRL_2006,Alberi_PRB_2007} and theoretical \cite{Usman_PRB_2011,Broderick_SST_2013} results. Comparing each of the GaBi$_{x}$As$_{1-x}$ PL features to the observed sharp features at 1.42 eV (associated with emission from the GaAs substrate) we see that the spectral width of the band edge PL increases with increasing Bi composition. This increase in the PL linewidth is primarily a result of the effects of alloy disorder, which causes strong inhomogeneous broadening of the VB edge states in GaBi$_{x}$As$_{1-x}$. \cite{Usman_PRB_2013}

The variation of the band gap and spin-orbit-splitting energy with Bi composition are crucial in determining the magnitude of $g^{*}$. In order to determine \DeltaSO\, precisely steady-state PR measurements were performed on all of the samples. The results of the PR measurements are summarized in Fig.~\ref{fig:energy_gaps}, which shows the measured variation of \Eg\, and \DeltaSO\, with $x$. Also shown in Fig.~\ref{fig:energy_gaps} are the results of large supercell tight-binding calculations of \Eg\, and \DeltaSO, which were performed using the theoretical model outlined in Section~\ref{sec:theoretical_model}. We note that the theoretical model accurately captures the strong, composition dependent bowing of \Eg\, and \DeltaSO, and is in good agreement with the experimental data across the investigated composition range.


\subsection{Measurement of $g^{*}$ via time-resolved photoluminescence spectroscopy}
\label{sec:measurement_of_g}

As a result of the $T_{d}$ point group symmetry present in unstrained bulk III-V compounds such as GaAs, $g^{*}$ is isotropic and independent of the direction of the applied magnetic field. \cite{Kalevich_JETPL_1993} However, in the presence of a reduction in symmetry, caused for example by quantum-confinement effects \cite{LeJeune_SST_1997,Malinowski_PRB_2000,Ivchenko_SPS_1992} or in the presence of pseudomorphic strain, \cite{Hendorfer_SST_1991} $g^{*}$ becomes anisotropic. Since the GaBi$_{x}$As$_{1-x}$ epilayers under investigation were grown on GaAs substrates, they are in a state of biaxial compressive stress. As a result of the increase in the GaBi$_{x}$As$_{1-x}$ lattice constant with increasing $x$, the difference between the transverse and longitudinal components of $g^{*}$ is expected to increase with increasing Bi composition.

Polarization-resolved PL spectroscopy can be used both to determine the magnitude of $g^{*}$, as well as to investigate its anisotropy. According to the optical selection rules in bulk material, when the pump beam propagates with circular polarization along the growth ($z$-) direction, the electron spins also align along $z$. Assuming rapid spin dephasing of holes, \cite{Sham_JPCM_1993,Hilton_PRL_2002,Meier_optical} the degree of electron spin polarization is obtained from the circular polarization $P_{c}$ of the luminescence

\begin{equation}
	P_{c} = \frac{ I^{+} - I^{-} }{ I^{+} + I^{-} } \, ,
\end{equation}

\noindent
where $I^{+}$ and $I^{-}$ are, respectively, the intensities of the co- and counter-polarized (with respect to the polarization of the exciting beam) PL components.


\begin{figure}[t!]
	\includegraphics[width=1.00\columnwidth]{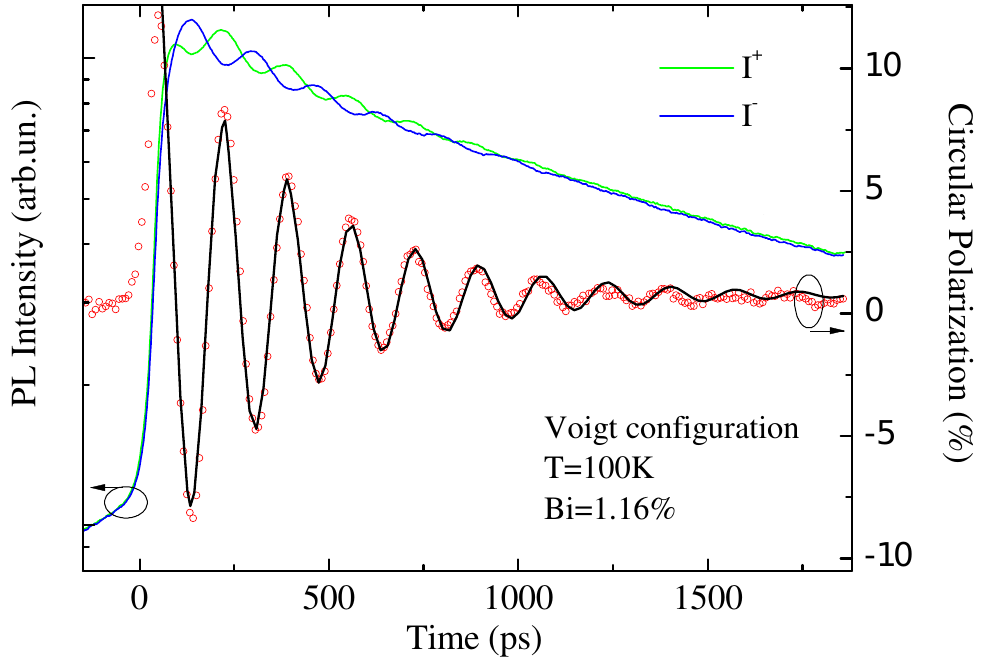}
	\caption{Green and blue lines, left $y$-axis: Temporal evolution of the co- and counter-polarized photoluminescence components of sample 1, measured in the Voigt configuration at $T = 100$ K with $B = 0.73$ T. Open red circles and black line, right $y$-axis: Temporal evolution of the luminescence circular polarization $P_{c}$ and an exponentially damped cosinusoidal fit to $P_{c}$ based on Eq.~\eqref{eq:spin_dynamics_fit}, respectively.}
	\label{fig:time_resolved_pl_1}
\end{figure}

When an external magnetic field is applied in the Voigt configuration, Larmor precession of spins around \textbf{B} occurs, giving rise to a quantum beating of the electron spins and thus of $P_{c}$. This behaviour is clearly visible in Fig.~\ref{fig:time_resolved_pl_1}, which shows the measured temporal evolution of the co- and counter-polarized PL components of sample 1 ($x = 1.16$\%) at $T = 100$ K, in the presence of a transverse magnetic field of 0.73 T. The measured temporal evolution of the degree of PL circular polarization $P_{c}$ is also shown in Fig.~\ref{fig:time_resolved_pl_1}. The latter is well described by assuming an exponential spin dephasing multiplied by a cosinusoidal oscillation describing the Larmor precession of electron spins about \textbf{B} at angular frequency $\omega$

\begin{equation}
	P_{c} (t) = P_{0} \, e^{ - t/\tau } \cos \left( \omega t + \phi\right) \, ,
	\label{eq:spin_dynamics_fit}
\end{equation}

\noindent
where $\tau$ is the electron spin dephasing time, $P_{0}$ is the initial degree of PL circular polarization, and $\phi$ is a constant phase shift.


\begin{figure}[t!]
	\includegraphics[width=1.00\columnwidth]{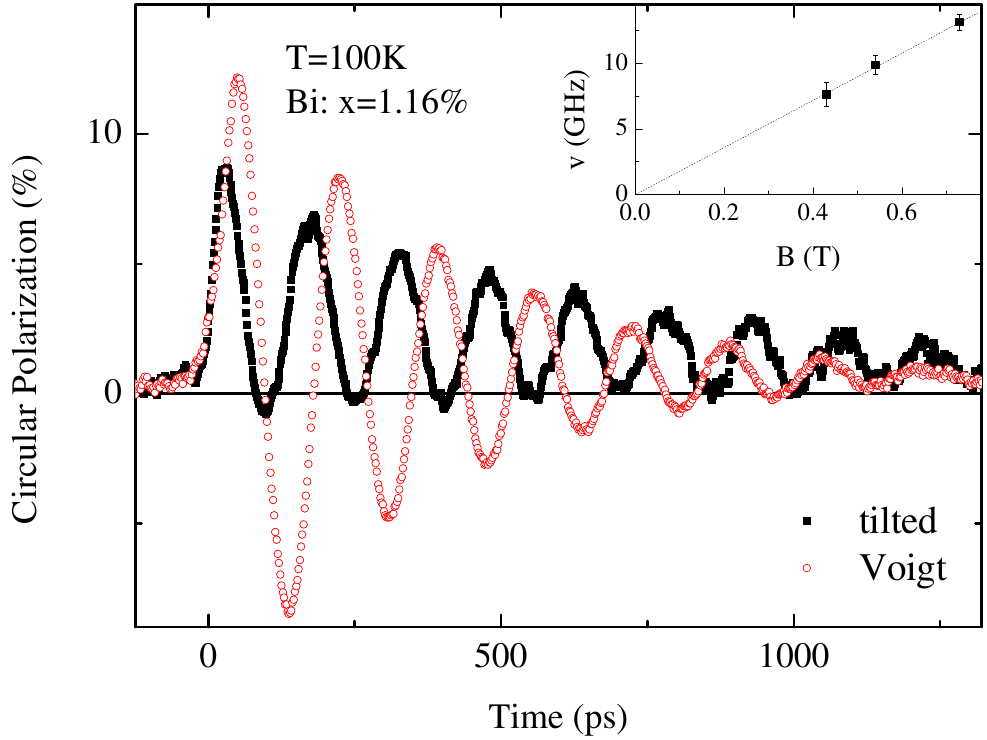}
	\caption{Comparison of the temporal evolution of the photoluminescence polarization of sample 1, measured in the transverse (Voigt; red open circles) and 45$^{\circ}$ (tilted; black closed circles) configurations at $T = 100$ K with $B = 0.73$ T. Inset: Measured variation of the Larmor frequency $\nu$ ($= \frac{\omega}{2 \pi}$) with the flux density $B$ of the externally applied magnetic field.}
	\label{fig:time_resolved_pl_3}
\end{figure}

The Larmor angular frequency of the electron spin precession is linked to the g factor tensor and applied field by

\begin{equation}
	\omega = \frac{ \mu_{\scalebox{0.6}{\textrm{B}}} B }{ \hbar } \, g^{*} \, ,
	\label{eq:larmor_frequency}
\end{equation}

\noindent
where $\mu_{\scalebox{0.6}{\textrm{B}}}$ is the Bohr magneton.

Experimentally, $\omega$ is obtained via analysis of the time-resolved PL polarization, which allows the magnitude of the relevant component of $g^{*}$ to be determined using Eq.~\eqref{eq:larmor_frequency} for a given relative orientation of the epilayer and applied magnetic field. In Ref.~\onlinecite{Mazzucato_APL_2013} we considered only the transverse component $g_{\perp}^{*}$ of the electron g factor, obtained by measuring $\omega$ in the Voigt configuration. Since there is no Larmor precession of electron spins when the applied magnetic field is parallel to the growth direction, determination of the longitudinal g factor $g_{\parallel}^{*}$ must be undertaken in a tilted configuration in which the applied magnetic field has two components, one in the layer plane as in the Voigt configuration, as well as an additional component along the growth ($z$-) direction. By orienting the epilayer at 45$^{\circ}$ with respect to the orientation of \textbf{B}, the magnitude of $g_{\parallel}^{*}$ can be determined via

\begin{equation}
	g_{\parallel}^{*} = \sqrt{ 2 g_{45^{\circ}}^{* \, 2} - g_{\perp}^{* \, 2} } \, ,
	\label{eq:g_factor_tilted_configuration}
\end{equation}

\noindent
where $g_{45^{\circ}}^{*}$ is the g factor determined in the 45$^{\circ}$ (tilted) configuration.\cite {LeJeune_SST_1997} Full details of the analysis leading to Eq.~\eqref{eq:g_factor_tilted_configuration} are given in the Appendix.


\begin{figure}[t!]
	\includegraphics[width=1.00\columnwidth]{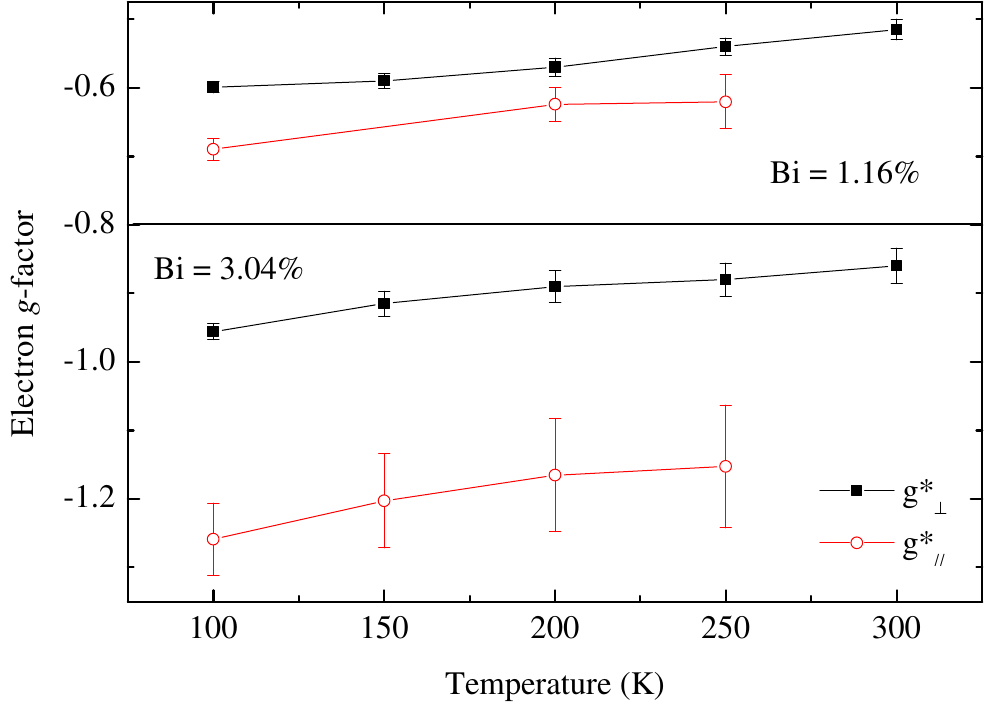}
	\caption{Measured variation of the transverse (black closed squares) and longitudinal (red open circles) components of the electron g factor with temperature in sample 1  ($x = 1.16$\%, top panel) and sample 4 ($x = 3.04$\%, bottom panel) .}
	\label{fig:g_factor_vs_temperature}
\end{figure}

A comparison of the measured polarization dynamics of sample 1 at $T = 100$ K in the Voigt and tilted configurations is shown in Fig.~\ref{fig:time_resolved_pl_3}, for an applied magnetic field with $B = 0.73$ T. We note that the Larmor frequency of the observed quantum beats is different in the two cases, with $\omega_{45^{\circ}} > \omega_{\perp}$, which indicates that $g^{*}$ is anisotropic and also that $\vert g_{\parallel}^{*} \vert > \vert g_{\perp}^{*} \vert$. The inset of Fig.~\ref{fig:time_resolved_pl_3} shows the measured variation of the quantum beating (Larmor) frequency as a function of the flux density of the applied field in the tilted configuration, confirming that $\omega$ increases linearly with $B$. By performing a linear fit to the variation of the measured Larmor angular frequency as a function of the magnetic flux density, $g_{45^{\circ}}^{*}$ can be determined from the slope of  $\omega_{45^{\circ}} (B)$ as per Eq.~\eqref{eq:larmor_frequency}.

Both $g_{\perp}^{*}$ and $g_{\parallel}^{*}$ increase strongly in magnitude with increasing Bi composition $x$, departing from the isotropic GaAs Land\'{e} g factor $g^{*} = -0.44$. \cite{Oestreich_PRL_1995} The measured values of $g_{\perp}^{*}$ and $g_{\parallel}^{*}$ for sample 1 ($x = 1.16$\%) and sample 4 ($x = 3.04$\%) are plotted as a function of temperature in Fig.~\ref{fig:g_factor_vs_temperature}. A reduction in the magnitude of both $g_{\perp}^{*}$ and $g_{\parallel}^{*}$ is observed with increasing temperature, which is consistent with previously observed trends in GaAs. \cite{Zawadzki_PRB_2008,Oestreich_PRL_1995,Oestreich_PRB_1996} We observe a larger difference between  $g_{\perp}^{*}$ and $g_{\parallel}^{*}$ in sample 4 than in sample 1 due to the fact that the pseudomorphic strain in the epilayer increases with increasing Bi composition $x$, leading to stronger anisotropy in $g^{*}$. Detailed explanation of this behaviour, as well as the dependence of $g^{*}$ on $x$, is provided by the results of our theoretical calculations in Section~\ref{sec:theoretical_results}. Our justification for the choice of a negative sign for the transverse and longitudinal components of $g^{*}$ in GaBi$_{x}$As$_{1-x}$ is based upon our previous measurement of $g_{\perp}^{*}$ in Ref.~\onlinecite{Mazzucato_APL_2013}, as well as the fact that $g^{*}$ is known to be negative in GaAs. \cite{Hermann_PRB_1977}

We note that the uncertainty associated with the measurement of $g^{*}$ increases with increasing temperature. The decrease in the magnitudes of $g_{\perp}^{*}$ and $g_{\parallel}^{*}$ with increasing temperature results in an increase in the period of the Larmor precession, which becomes comparable to the electron spin dephasing time at higher temperatures. Rapid decrease of the spin dephasing time has been observed not only with increasing temperature, but also with increasing Bi composition $x$ in GaBi$_{x}$As$_{1-x}$, as expected based on the theoretical analysis of Ref.~\onlinecite{Tong_JAP_2012}. The latter is a result of the strong enhancement of the spin-orbit coupling in GaBi$_{x}$As$_{1-x}$\, with increasing $x$. Spin dephasing therefore reduces the accuracy with which the quantum beating frequency, and hence $g^{*}$, can be measured at high temperature. We further note that, at all temperatures and Bi compositions, the uncertainty associated with the measurement of $g_{\parallel}^{*}$ is greater than that related to $g_{\perp}^{*}$ due to the increased noise present in measurements of the time-resolved PL in the tilted configuration compared to the Voigt configuration. This increased noise in the PL measured in the tilted configuration, combined with dephasing of the electron spins on a timescale comparable to the Larmor period at high temperature, meant that it was not possible to measure $g_{\parallel}^{*}$ at 300 K. Similarly, the increased uncertainty associated with the measured values of $g_{\perp}^{*}$ and $g_{\parallel}^{*}$ at higher $x$ is due to the Bi-induced reduction of the spin dephasing time.


\subsection{Theoretical results and comparison with experimental measurements of $g^{*}$ }
\label{sec:theoretical_results}

In order to investigate the effects of alloy disorder and pseudomorphic strain on the anisotropy of the electron g factor, we have used the theoretical model described in Section~\ref{sec:theoretical_model} to calculate the variation of the transverse and longitudinal components of $g^{*}$ as a function of $x$. The theoretical calculations were performed on a series of disordered 4096-atom supercells, which are either free-standing (unstrained) or under compressive pseudomorphic strain. In the free-standing supercells all atomic positions are allowed to relax freely, with no constraints applied to the atoms at the supercell boundaries. In order to directly quantify the effects of pseudomorphic strain on the electron g factor, the same spatial distributions of Bi atoms were used at each composition $x$ in the unstrained and strained supercell calculations.


\begin{figure}[t!]
	\includegraphics[width=1.00\columnwidth]{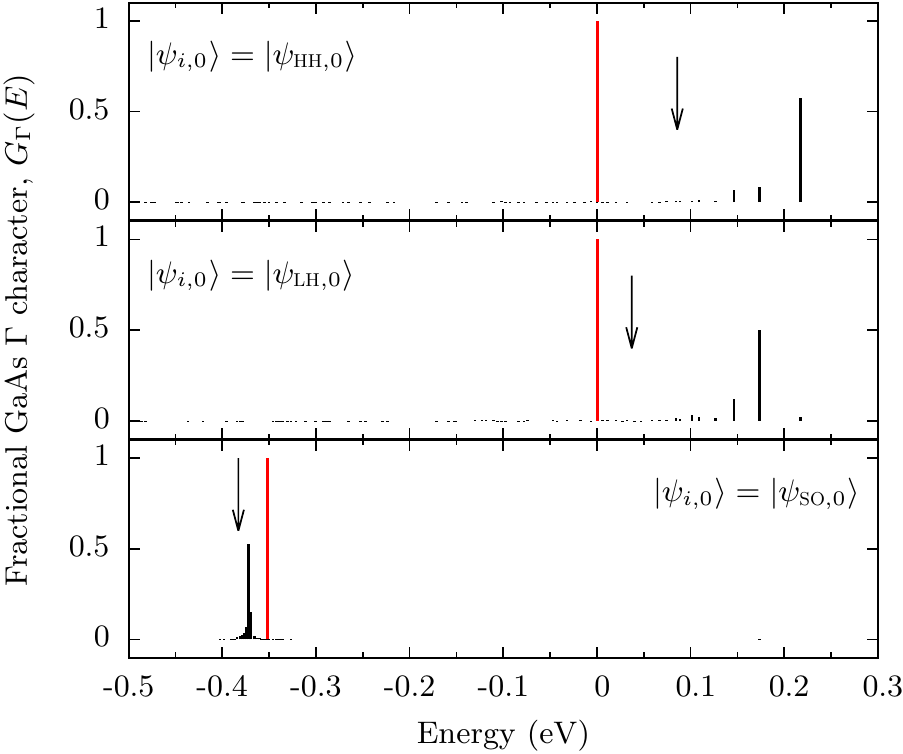}
	\caption{Black lines: Calculated distribution of the $\Gamma$ character associated with the GaAs heavy-hole (top panel), light-hole (middle panel) and spin-split-off-hole (bottom panel) band edge states over the full spectrum of alloy zone center valence states in a disordered, 4096-atom Ga$_{2048}$Bi$_{82}$As$_{1966}$ ($x = 4$\%) supercell. The supercell was placed under a compressive pseudomorphic strain, corresponding to epitaxial growth on a GaAs substrate. Red lines show the calculated distribution of $\Gamma$ character in a Bi-free Ga$_{2048}$As$_{2048}$ supercell. The zero of energy is taken at the GaAs valence band maximum. The black vertical arrows in each panel denote the weighted average energies of the HH-, LH- and SO-like alloy valence states, calculated using Eq.~\eqref{eq:replacement} for each of $E_{A}$, $E_{B}$ and $E_{C}$.}
	\label{fig:gamma_character}
\end{figure}

Due to the fact that the tight-binding model which we use underestimates the magnitude of the inter-band momentum matrix element $P$ by approximately 25\%, \cite{Broderick_SST_2013} and since we neglect the effects of the $\Gamma_{7c}$ and $\Gamma_{8c}$ CB states in our calculations of $g^{*}$, we obtain a suitable value for $P$ by fitting to the experimental value of $g^{*}$ in GaAs, where the values of \Eg\, and \DeltaSO\, are taken directly from a tight-binding calculation with $x = 0$. Following this procedure we obtain a value of $P = 10.13$ eV \AA\, for GaAs at $T = 300$ K, close to the commonly accepted value of 10.48 eV \AA. \cite{Vurgaftman_JAP_2001} This value of $P$ is then kept constant in all calculations of $g_{\perp}^{*}$ and $g_{\parallel}^{*}$ as the Bi composition $x$ is varied. Based on the analysis of Ref.~\onlinecite{Zawadzki_PRB_2008}, the variation of the GaAs host matrix band gap with temperature is taken to depend solely on the effects of lattice dilatation.

Fig.~\ref{fig:gamma_character} shows the calculated distribution of the $\Gamma$ character associated with the extended GaAs host matrix HH, LH and SO band edge states over the full spectrum of zone center alloy states, for a disordered 4096-atom Ga$_{2048}$Bi$_{82}$As$_{1966}$ ($x = 4$\%) supercell under compressive pseudomorphic strain. The top, middle and bottom panels of Fig.~\ref{fig:gamma_character} show, respectively, the $G_{\Gamma}^{\scalebox{0.6}{\textrm{HH}}} \left( E_{j,1} \right)$, $G_{\Gamma}^{\scalebox{0.6}{\textrm{LH}}} \left( E_{j,1} \right)$ and $G_{\Gamma}^{\scalebox{0.6}{\textrm{SO}}} \left( E_{j,1} \right)$ spectra calculated using Eq.~\eqref{eq:gamma_character}. For comparative purposes, the corresponding $\Gamma$ character of the unperturbed GaAs host matrix is also shown in Fig.~\ref{fig:gamma_character}. As is expected for an epilayer under compressive strain, we see that the topmost alloy VB state is predominantly HH-like, having $> 50$\% GaAs HH $\Gamma$ character. We additionally note that both the LH and HH band edges in the alloy are strongly inhomogeneously broadened, which is evidenced by the large number of alloy VB states over which the GaAs VB edge $\Gamma$ character is distributed.

We have previously demonstrated that the strong inhomogeneous broadening of the GaBi$_{x}$As$_{1-x}$ VB edge states is a result of (i) the strong broadening of localized states associated with isolated Bi atoms due to the large density of GaAs VB states with which they are resonant, \cite{Broderick_SST_2013,Fano_PR_1961} and (ii) the formation of a range of localized states associated with Bi-Bi pairs and larger clusters of Bi atoms, which lie above the GaAs VB edge in energy and hybridize strongly with the alloy VB edge states. \cite{Usman_PRB_2013} In contrast to the character of the alloy VB edge, we note that the feature associated with the $\Gamma$ character of the GaAs SO band edge states, visible in the bottom panel of Fig.~\ref{fig:gamma_character}, remains relatively sharp in GaBi$_{x}$As$_{1-x}$. The calculated distribution of GaAs SO character shows that the strong inhomogeneous broadening of the LH-like alloy VB states minimizes the expected strain-induced mixing between the LH- and SO-like states. Further calculations that we have undertaken show strain-induced mixing between the LH and SO states in ordered supercells; the strong inhibition of such mixing observed in Fig.~\ref{fig:gamma_character} is therefore due to the disorder-induced broadening of the GaAs LH states in the disordered alloy supercells.

\begin{figure}[t!]
	\includegraphics[width=1.00\columnwidth]{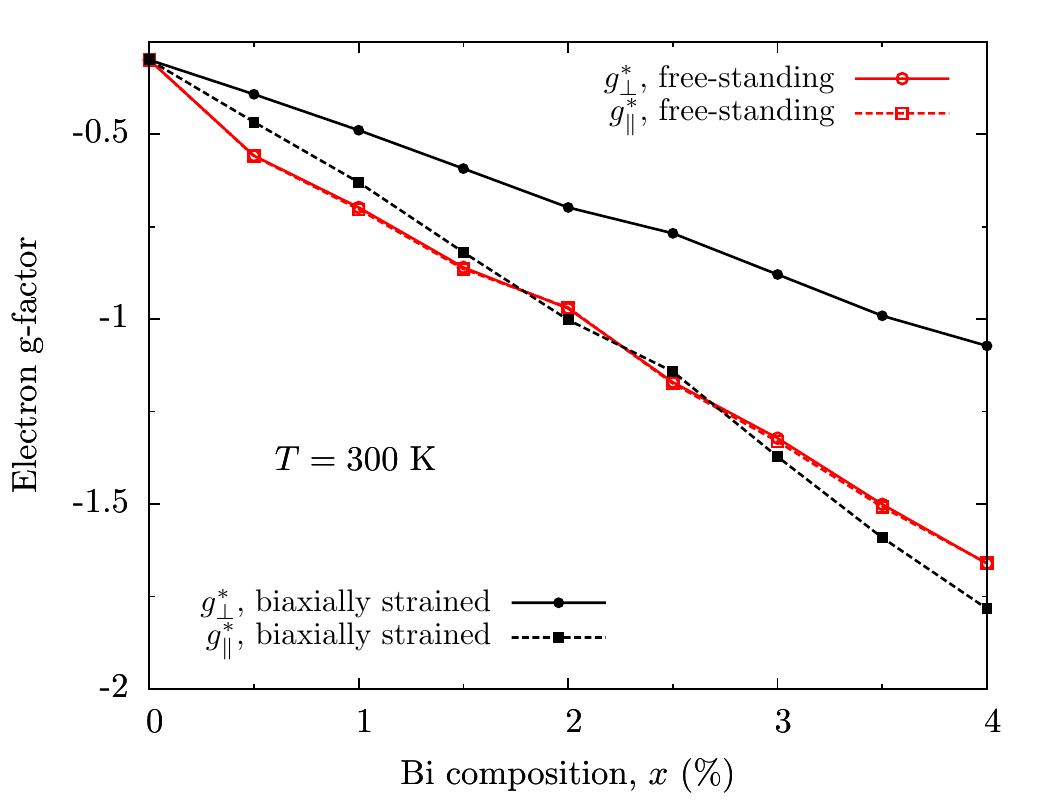}
	\caption{Calculated variation of the transverse and longitudinal components of the electron g factor (denoted respectively using circles and squares) at $T = 300$ K, as a function of Bi composition $x$. Red open symbols denote free-standing calculations in which no macroscopic strain has been applied to the supercells. Black closed symbols denote calculations in which a compressive pseudomorphic strain corresponding to epitaxial growth on a GaAs substrate has been applied to the supercells.}
	\label{fig:g_free_standing_comparison_300K}
\end{figure}

In the absence of pseudomorphic strain $g^{*}$ is expected to be isotropic, with $g_{\perp}^{*} = g_{\parallel}^{*}$. \cite{Hendorfer_SST_1991} Examining the results of our free-standing supercell calculations in Fig.~\ref{fig:g_free_standing_comparison_300K} we see that this isotropy holds in GaBi$_{x}$As$_{1-x}$ alloys, even in the presence of strong alloy disorder. In addition to a rapid, monotonic increase in the magnitude of $g^{*}$ with increasing $x$, we also calculate that the reduced local symmetry present in the disordered alloy supercells has a negligible effect on the isotropy of $g^{*}$. In all of the supercells considered the calculated values of $g_{\perp}^{*}$ and $g_{\parallel}^{*}$ are approximately equal, even in the presence of significant alloy disorder at $x = 4$\%. For example, the maximum calculated difference between the magnitudes of $g_{\perp}^{*}$ and $g_{\parallel}^{*}$ in an individual disordered supercell at $x = 4$\% was 0.034, representing a disorder-induced variation of approximately 2\% in the magnitude of $g^{*}$, so that the g factor can be considered effectively isotropic even in strongly disordered GaBi$_{x}$As$_{1-x}$ alloys.

Further examining Fig.~\ref{fig:g_free_standing_comparison_300K} we see that this situation changes drastically in the presence of pseudomorphic strain, in which a strong anisotropy of the g factor appears so that $g_{\perp}^{*} \neq g_{\parallel}^{*}$. As in the unstrained case we see that the magnitudes of both $g_{\perp}^{*}$ and $g_{\parallel}^{*}$ increase strongly and monotonically with increasing $x$. We also calculate that (i) the magnitude of $g_{\parallel}^{*}$ exceeds that of $g_{\perp}^{*}$ for non-zero $x$, in agreement with experiment, and (ii) the magnitude of $g_{\parallel}^{*}$ is close to the isotropic value of $g^{*}$ calculated in the absence of strain, both of which are expected for epitaxial layers under compressive strain. \cite{Hendorfer_SST_1991} Based on these comparative calculations we conclude that while the symmetry of $g^{*}$ is essentially unaltered by the effects of alloy disorder, the effects of pseudomorphic strain are crucial for determining the experimentally observed anisotropy of $g^{*}$ in GaBi$_{x}$As$_{1-x}$ epilayers, as is generally expected for III-V alloys. \cite{Hendorfer_SST_1991}

\begin{figure}[t!]
	\includegraphics[width=1.00\columnwidth]{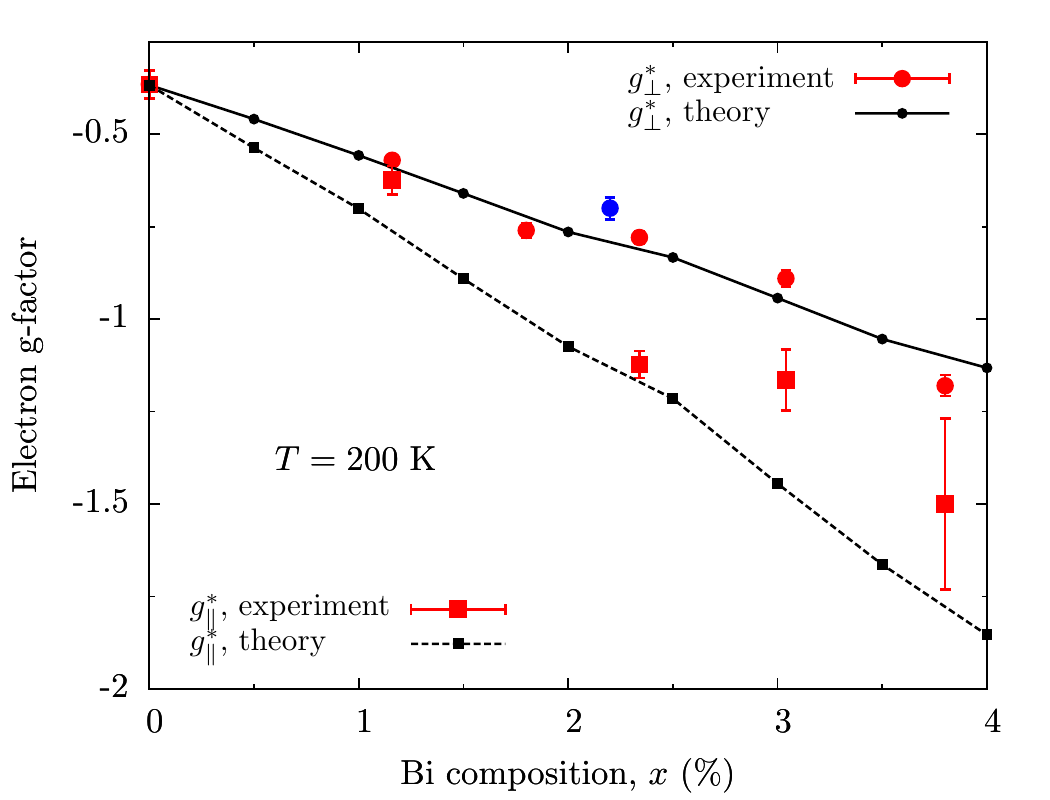}
	\caption{Comparison between the measured and calculated variation of the transverse and longitudinal components of the electron g factor at $T = 200$ K, as a function of Bi composition $x$ in a series of pseudomorphically strained GaBi$_{x}$As$_{1-x}$/GaAs epilayers. Red closed circles (squares) and black closed circles (squares) denote the measured and calculated values of $g_{\perp}^{*}$ ($g_{\parallel}^{*}$), respectively. The measured value of $g_{\perp}^{*}$ at $x = 2.2$\% from Ref.~\onlinecite{Mazzucato_APL_2013} is shown using a blue closed circle.}
	\label{fig:g_factor_theory_vs_experiment}
\end{figure}

Fig.~\ref{fig:g_factor_theory_vs_experiment} shows the measured and calculated variation of $g_{\perp}^{*}$ and $g_{\parallel}^{*}$ with $x$ at $T = 200$ K, where the theoretical calculations include the appropriate composition dependent pseudomorphic strain. In addition to the experimental data for samples 1 -- 5, the measured value of $g_{\perp}^{*}$ at $x = 2.2$\% has been included from Ref.~\onlinecite{Mazzucato_APL_2013} for completeness. We see that the theoretical model reproduces the observed enhancement of $g_{\perp}^{*}$ with increasing $x$ to a high degree of accuracy, but that the description of $g_{\parallel}^{*}$ is comparatively less accurate. One possible source of the observed discrepancy between the measured and calculated values of the longitudinal component of the g factor is the difficulty related to obtaining a suitably strong PL signal for the determination of the Larmor frequency in the 45$^{\circ}$ tilted configuration. This is reflected in the large experimental errors associated with the measured values of $g_{\parallel}^{*}$, as discussed in Section~\ref{sec:measurement_of_g}. Additionally, reciprocal space maps measured around the asymmetric (224)+ direction did not show any evidence of plastic relaxation in the epilayers. However, transmission electron diffraction measurements performed on sample 5 ($x = 3.83$\%) showed the presence of a low density of interfacial dislocations, which indicates that plastic relaxation has started to occur. The onset of plastic relaxation in samples with higher Bi compositions may also partially explain the observed discrepancy between the measured and calculated values of $g_{\parallel}^{*}$.


In Ref.~\onlinecite{Mazzucato_APL_2013} it was shown that using Eq.~\eqref{eq:g_factor_hendorfer_per} to calculate the variation of $g_{\perp}^{*}$ with $x$, by setting $E_{B} = E_{g} (x)$ and $E_{C} = E_{g} (x) + \Delta_{\scalebox{0.6}{\textrm{SO}}} (x)$, strongly overestimates the experimentally observed increase in the magnitude of $g_{\perp}^{*}$ with increasing $x$. The failure of this simple approach can be understood in terms of the results presented in Fig.~\ref{fig:gamma_character}, which show that (i) the separation in energy between the CB and LH band edges is not equal to $E_{g} (x)$ due to the strain- and disorder-induced splitting of the alloy VB edge, and (ii) the weighted average of the LH contributions to $g^{*}$ lie at lower energies in the VB of the disordered alloy, due to the strong, Bi-induced hybridization of the alloy VB states.

In the theoretical model of Section~\ref{sec:theoretical_model}, the effects of alloy disorder on the calculated variation of $g_{\perp}^{*}$ and $g_{\parallel}^{*}$ with $x$ can be determined by considering the distribution of host matrix $\Gamma$ character shown in Fig.~\ref{fig:gamma_character}. In the disordered alloy the VB edge GaAs $\Gamma$ character is distributed over a large number of alloy VB states, with the details of the distribution determined by the precise disorder present in the alloy. The details of this distribution of the GaAs $\Gamma$ character onto VB states lying below the alloy VB edge therefore plays a crucial role in determining the calculated variation of $g_{\perp}^{*}$ and $g_{\parallel}^{*}$ with $x$ since, by Eqs.~\eqref{eq:g_factor_per} and~\eqref{eq:g_factor_par}, the contribution from a given alloy VB state to a specific component of $g^{*}$ is weighted by its associated host matrix $\Gamma$ character. Specifically, mixing of this $\Gamma$ character into VB states with greater separation in energy from the alloy CB minimum results in contributions from terms such as those in Eq.~\eqref{eq:replacement} with larger denominators, the net result of which is to act against the stronger increase in $g_{\perp}^{*}$ and $g_{\parallel}^{*}$ that would be expected based solely upon the measured variation of \Eg\, and \DeltaSO\, with $x$.


To further illustrate this point within the context of our theoretical model we have used the values of $g_{\perp}^{*}$ and $g_{\parallel}^{*}$ calculated using Eqs.~\eqref{eq:g_factor_per} and~\eqref{eq:g_factor_par} at $T = 200$ K for the supercell of Fig.~\ref{fig:gamma_character}, as well as the weighted average energy of the sharp SO-related feature calculated using the GaAs SO character spectrum (denoted by a vertical arrow in the bottom panel of Fig.~\ref{fig:gamma_character}), to solve for the values of $E_{A}$ and $E_{B}$ in Eqs.~\eqref{eq:g_factor_hendorfer_per} and~\eqref{eq:g_factor_hendorfer_par} that would be required to reproduce these same values of $g_{\perp}^{*}$ and $g_{\parallel}^{*}$. The vertical arrows in the top and middle panels of Fig.~\ref{fig:gamma_character} show the HH- and LH-like band edge energies obtained from the values of $E_{A}$ and $E_{B}$ calculated in this manner (equivalent to calculating $E_{A}$ from Eq.~\eqref{eq:replacement}, with a similar expression for $E_{B}$ in terms of $G_{\Gamma}^{\scalebox{0.6}{\textrm{LH}}} \left( E_{j,1} \right)$). Examining the calculated distribution of GaAs HH and LH character in the disordered supercell, we see that the values of $E_{A}$ and $E_{B}$ required to reproduce $g_{\perp}^{*}$ and $g_{\parallel}^{*}$ as obtained from Eqs.~\eqref{eq:g_factor_per} and~\eqref{eq:g_factor_par} lie at significantly lower energy than the alloy VB states which account for the majority of the GaAs VB edge character. As a result, simple calculations based on Eqs.~\eqref{eq:g_factor_hendorfer_per} and~\eqref{eq:g_factor_hendorfer_par} that do not take into account the strong, Bi-induced mixing of VB states in a disordered GaBi$_{x}$As$_{1-x}$ alloy necessarily break down unless unrealistic values of $E_{A}$ and $E_{B}$ are used to represent the energies of the alloy VB edge states. We conclude that alloy disorder therefore plays a key role in determining the variation of the magnitudes of $g_{\perp}^{*}$ and $g_{\parallel}^{*}$ with $x$ in GaBi$_{x}$As$_{1-x}$. Based on this analysis we conclude that pseudomorphic strain plays a key role in determining the magnitudes of $g_{\perp}^{*}$ and $g_{\parallel}^{*}$ at a given Bi composition in the disordered alloy since, when compared to the equivalent unstrained material, compressive strain acts to preferentially redistribute the GaAs HH character onto higher energy alloy VB states.

We note that the theoretical calculations of $g^{*}$ presented here, as well as further calculations that we have undertaken, indicate a monotonic decrease of the electron effective mass at the CB minimum ($m_{c}^{*}$) with increasing $x$. While this is in agreement with the experimental measurements presented in Ref.~\onlinecite{Fluegel_APL_2011}, it contradicts those presented in Ref.~\onlinecite{Pettinari_PRB_2010} which identified an unexpected increase of $m_{c}^{*}$ with increasing $x$. Therefore, while conflicting sets of experimental data exist on the evolution of $m_{c}^{*}$ with $x$ in GaBi$_{x}$As$_{1-x}$ alloys and there appears as yet to be no consensus in the literature, our theoretical analysis supports those experimental data that indicate a decrease in $m_{c}^{*}$ with increasing $x$. However, we note that the $\Gamma_{7c}$ and $\Gamma_{8c}$ CB states (the effects of which are neglected in our current analysis) play a significantly more important role in determining $m_{c}^{*}$ than they do in determining $g^{*}$. \cite{Hermann_PRB_1977} As a result, a detailed discussion of the effects of Bi incorporation on $m_{c}^{*}$ is beyond the scope of the current work.

Overall, the theoretical model we developed in Section~\ref{sec:theoretical_model} is capable of describing the anisotropy of $g^{*}$ in pseudomorphically strained GaBi$_{x}$As$_{1-x}$ epilayers, predicting correctly that $\vert g_{\parallel}^{*} \vert > \vert g_{\perp}^{*} \vert$ in addition to describing the observed strong enhancement of the magnitudes of $g_{\perp}^{*}$ and $g_{\parallel}^{*}$ with increasing $x$. We conclude that the theoretical calculations are in very good agreement with the observed experimental trends, which reinforces the validity of the detailed picture of the electronic structure of disordered GaBi$_{x}$As$_{1-x}$ alloys provided by the tight-binding model. \cite{Usman_PRB_2011,Usman_PRB_2013}


\section{Conclusions}
\label{sec:conclusions}

We have presented a detailed experimental and theoretical study of the composition dependence of the electron Land\'{e} g factor $g^{*}$ in GaBi$_{x}$As$_{1-x}$/GaAs epilayers, for Bi compositions up to $x = 3.8$\%. The optical properties of these epilayers were characterized using a combination of time-integrated photoluminescence and photo-modulated reflectance measurements. The observed rapid reduction of the GaBi$_{x}$As$_{1-x}$ band gap, and increase in the spin-orbit-splitting energy, with increasing $x$ is in line with that noted previously by several authors. 

The transverse and longitudinal components of $g^{*}$ were measured via time-resolved photoluminescence spectroscopy, which was used to determine the frequency of the quantum beating of the luminescence circular polarization associated with the Larmor precession of electron spins about an externally applied magnetic field. By varying the orientation of the epilayer samples with respect to the direction of the applied magnetic field we demonstrated that (i) $g^{*}$ becomes highly anisotropic in pseudomorphically strained GaBi$_{x}$As$_{1-x}$ alloys, and (ii) the magnitudes of the transverse ($g_{\perp}^{*}$) and longitudinal ($g_{\parallel}^{*}$) components of $g^{*}$ increase strongly with increasing Bi composition $x$.

In order to explain this behaviour we have developed a theoretical model for the description of $g^{*}$ in GaBi$_{x}$As$_{1-x}$ alloys. The theoretical model is based upon atomistic tight-binding calculations of the electronic structure of large GaBi$_{x}$As$_{1-x}$ alloy supercells and explicitly takes account of the effects of alloy disorder and pseudomorphic strain, which are known to play key roles in determining the details of the valence band structure in dilute bismide alloys. Calculations undertaken using this model have revealed that the rapid increase in the magnitude of both $g_{\perp}^{*}$ and $g_{\parallel}^{*}$ with increasing $x$ in GaBi$_{x}$As$_{1-x}$ is explained by a combination of the strong decrease (increase) in the band gap (spin-orbit-splitting energy) with increasing $x$, as well as hybridization of the extended states of the GaAs host matrix valence band edge with the large range of localized Bi-related states present in a disordered GaBi$_{x}$As$_{1-x}$ alloy.

Our theoretical analysis has additionally revealed that the reduced symmetry in strongly disordered GaBi$_{x}$As$_{1-x}$ alloys has a negligible effect on the isotropy of $g^{*}$. The observed strong anisotropy of $g^{*}$ in the bulk-like epitaxial layers studied here is then attributed to the effects of compressive strain, as is the case for other III-V material systems.

Overall, our measured and calculated results are in very good agreement and our combined experimental and theoretical analysis has elucidated important factors concerning the strongly perturbative effects of Bi incorporation on the electronic structure GaBi$_{x}$As$_{1-x}$ alloys.


\section*{Acknowledgements}

CB, MU and EOR acknowledge support from the Irish Research Council (EMBARK Postgraduate Scholarship no.~RS/2010/2766 to CB), the European Union Seventh Framework Programme (BIANCHO; project no.~FP7-257974) and Science Foundation Ireland (project no.~10/IN.1/I299). OD and AE acknowledge support from the Scientific Research Projects Coordination Unit of Istanbul University (project nos.~27643, 31160 and 44649) and the Ministry of Development of the Turkish Republic (project no.~2010 K121050). LPCNO and LAAS members acknowledge support from Toulouse Tech InterLabs 2013, as well as the support of the LAAS technology platform (RENATECH) for facilitating the MBE growth.


\section*{Appendix}
\label{sec:appendix}

In order to directly determine the components of the electron g factor tensor $g^{*}$, non-resonant time-resolved PL measurements were carried out in the presence of an externally applied magnetic field. We describe here the theory underlying the procedure used to obtain the transverse and longitudinal components $g_{\perp}^{*}$ and $g_{\parallel}^{*}$ of the anisotropic g factor from the observed quantum beating of the circular polarization $P_{c}$ of PL from the GaBi$_{x}$As$_{1-x}$ epilayers.

The temporal evolution of the components of the electron spin $\textbf{S} = ( S_{x}, S_{y}, S_{z} )$, in the presence of a tilted magnetic field $\textbf{B} = ( B_{x}, 0, B_{z} )$, is given by \cite{LeJeune_SST_1997}

\begin{equation}
	\frac{ \textrm{d} S_{i} }{ \textrm{d} t } = \left( \textbf{$\omega$} \times \textbf{S} \right)_{i} - \frac{ S_{i} }{ \tau_{i} } \, ,
	\label{eq:spin_evolution}
\end{equation}

\noindent
where $i = x, y, z$ and the anisotropic spin dephasing times $\tau_{i}$ are given along the transverse and longitudinal directions by $\tau_{x} = \tau_{y} = \tau_{\perp}$ and $\tau_{z} = \tau_{\parallel}$ respectively. The vector \textbf{$\omega$} is the Larmor angular frequency of the gyrating electron spins, the components of which are given by

\begin{eqnarray}
	\omega_{x} &=& \frac{ \mu_{\scalebox{0.6}{\textrm{B}}} B_{x} }{ \hbar } \, g_{\perp}^{*} \, , \label{eq:larmor_frequency_x} \\
	\omega_{y} &=& 0 \, , \\
	\omega_{z} &=& \frac{ \mu_{\scalebox{0.6}{\textrm{B}}} B_{z} }{ \hbar } \, g_{\parallel}^{*} \, .
\end{eqnarray}

The first term on the right hand side of Eq.~\eqref{eq:spin_evolution} describes the coherent evolution resulting from the coupling of the electron spin \textbf{S} to the magnetic field \textbf{B}, while the second term describes the spin dephasing dynamics. Under experimental conditions in which \cite{Kalevich_PSS_1997}

\begin{equation}
	\omega \gg \frac{1}{2} \, \bigg| \frac{1}{\tau_{\perp}} - \frac{1}{\tau_{\parallel}} \bigg| \, ,
\end{equation}

\noindent
if we take an initial spin vector $\textbf{S}_{0} = ( 0, 0, S_{0})$ and ignore the effects of spin dephasing, then we calculate the $z$-component of the electron spin by solving Eq.~\eqref{eq:spin_evolution} to obtain

\begin{equation}
	S_{z} (t) =  S_{0} \left[ 1 - \left( \frac{ \omega_{x} }{ \omega } \right)^{2} \, \bigg( 1 - \cos \left( \omega t \right) \bigg) \right] \, ,
\end{equation}

\noindent
where $\omega = \sqrt{ \omega_{x}^{2} + \omega_{z}^{2} } \, $.

In Voigt configuration, when $\textbf{B}$ has only one component, along the $x$-axis, $S_{z} (t)$ is given by

\begin{equation}
	S_{z} (t) =  S_{0} \cos \left( \omega t  \right) \, .
\end{equation}

Experimentally, the Larmor angular frequency $\omega$ is extracted from the quantum beats of the circular polarization of the PL, $P_{c} (t) = - 2 S_{z} (t)$, as a function of the intensity of the applied magnetic field.

In the Voigt and 45$^{\circ}$ titled configurations -- in which the magnetic fields are directed at $90^{\circ}$ and $45^{\circ}$ to the growth ($z$-) direction respectively, with $\textbf{B} = ( B, 0, 0 )$ and $\textbf{B} = \left( \frac{B}{\sqrt{2}}, 0, \frac{B}{\sqrt{2}} \right)$ -- the Larmor angular frequencies of the quantum beats in the polarization of the PL are given by

\begin{eqnarray}
	\omega_{90^{\circ}} &=& \omega_{\perp} = \frac{ \mu_{\scalebox{0.6}{\textrm{B}}} B }{ \hbar } \, g_{\perp}^{*} \, , \\
	\omega_{45^{\circ}} &=& \frac{ \mu_{\scalebox{0.6}{\textrm{B}}} B }{ \hbar } \, g_{45^{\circ}}^{*} = \frac{ \mu_{\scalebox{0.6}{\textrm{B}}} B }{ \hbar } \sqrt{ \frac{ g_{\parallel}^{* \, 2} + g_{\perp}^{* \, 2} }{ 2 } } \, . \label{eq:g_factor_tilted}
\end{eqnarray}

Therefore, once the values of $g_{\perp}^{*}$ and $g_{45^{\circ}}^{*}$ are determined by measuring the frequency of the quantum beats in the PL circular polarization for each of the Voigt and $45^{\circ}$ tilted configurations, $g_{\parallel}^{*}$ is determined straightforwardly from Eq.~\eqref{eq:g_factor_tilted} as

\begin{equation}
	g_{\parallel}^{*} = \sqrt{ 2 g_{45^{\circ}}^{* \, 2} - g_{\perp}^{* \, 2} } \, , \nonumber
\end{equation}

\noindent
which is Eq.~\eqref{eq:g_factor_tilted_configuration} in the main text.



\end{document}